\documentclass[pre,showpacs,preprint]{revtex4}

\begin{document}

\title{Thermodynamics and the virial expansion for trapped fluids in arbitrary external potentials} 

\author{Nadia Sandoval-Figueroa and V\'{\i}ctor Romero-Roch\'{\i}n\footnote{Corresponding author}}

\email{nadia@fisica.unam.mx, romero@fisica.unam.mx}

\affiliation{Instituto de F\'{\i}sica, Universidad Nacional
Aut\'onoma de M\'exico. \\
Apartado Postal 20-364, 01000 M\'exico,
D.F. Mexico.}

\begin{abstract}

 We present the full thermodynamics of a fluid confined by an arbitrary external potential based on the virial expansion of the grand potential. The fluid may be classical or quantum and it is assumed that interatomic interactions are pairwise additive. We indicate how the appropriate ``generalized" volume and pressure variables, that replace the usual volume and hydrostatic pressure, emerge for a given confining potential in the thermodynamic limit. A discussion of the physical meaning and of the measurement of these variables is presented. We emphasize that this treatment yields the correct equation of state of the fluid and we give its virial expansion. We propose an experiment to measure the heat capacity, so that with this quantity and the equation of state, the complete thermodynamics of the system may be extracted. As a corollary, we find that the so-called {\it local density approximation}  for these systems follows in the thermodynamic limit, although we also point out that it cannot be used indiscriminately for all local variables. Along the text we discuss the relevance of these findings in the description of the currently confined ultracold gases.
 
 \end{abstract}


\date{\today}
\maketitle

\section{Introduction.}

The intense experimental and theoretical activity in the field of 
quantum ultracold {\it trapped} 
gases\cite{Anderson,Davies,Bradley,Greiner,Bartenstein,Thomas1,Magalhaes,Ziewerlein,Hulet,Thomas2,Shin,Butts,Dalfovo,Perali,RR1,RR2,RR3,DeSilva,Gubbels,Chien,Ho,Drummond,Thomas3} has largely stimulated first principles analysis of the physics of these inhomogeneous systems.  In particular, 
the calculation of density profiles has turned out to be of fundamental
importance since it is one of the main measurable quantities in 
the recent experiments. From the density profile several thermodynamic quantities can be obtained, such as the number of particles and the temperature. However, because the hydrostatic pressure is a {\it local} quantity and the volume of the system cannot be rigorously defined, the use of  thermodynamics as a tool to analyze and characterize such trapped gases has been of limited application. For instance, important properties and quantities, such as the equation of state of the fluid or its heat capacities, are lacking. As we discuss here, those properties can actually be extracted from the knowledge of the density profile and other simple properties of the trapped gases.

 The main purpose of this article is to emphasize the fact that the thermodynamics of trapped systems must be reformulated in terms of the appropriate mechanical variables that, for lack of a better name, we shall call {\it generalized} pressure ${\cal P}$ and volume ${\cal V}$. That is, the usual hydrostatic pressure $p$ and the volume $V$ of a fluid contained in a vessel of rigid walls are no longer thermodynamic variables for a fluid confined by an external {\it inhomogeneous} field $V_{ext}(\vec r)$. As mentioned above, in the presence of an arbitrary confining potential, the pressure of the fluid becomes a local variable $p = p(\vec r)$ and the volume is strictly undefined. As we shall show there is a ``new'' and unique pair of variables ${\cal P}$ and ${\cal V}$ that replace the usual ones, $p$ and $V$. We shall illustrate these variables with specific examples. Although this result has already been pointed out for a gas trapped in a harmonic potential\cite{RR1,RR2,RR3}, its use has been limited within the approximations that the gas is ideal or at best that the interactions may be treated \`a la Hartree-Fock\cite{RR1}; nevertheless, incipient comparisons have been made with experiments in ultracold Na gases\cite{Silva,Henn,Henn2} showing its potential usefulness. In any case, we believe the main point has not been fully appreciated, namely, the fact that a different set of thermodynamics variables must be used for a given external confining potential. We proceed here to fill this gap, deducing the virial expansion for the Grand Potential in the grand canonical ensemble for an arbitrary external potential, valid for classical and quantum fluids, either ideal or with pairwise interatomic interactions. Extensions to three- or higher-body interactions may be further considered. We discuss the physical meaning of the generalized pressure and its relevance regarding the {\it equation of state} of the fluid,  i.e. ${\cal P} = {\cal P}({\cal V}/N,T)$ with $N$ the number of molecules or atoms in the fluid and $T$ the temperature. This finding should have immediate practical applications since all the thermodynamic properties, specially phase transitions, can be quantitatively described and certainly visualized in the corresponding phase diagram. We emphasize the fact that the generalized pressure can be very simply measured or calculated once the particle density profile $\rho(\vec r)$ is known.  This procedure has already been exploited in the analysis of experimental data of trapped Na ultracold gases in quadrupolar\cite{Silva} and in harmonic traps\cite{Henn,Henn2}. As a further step, we propose a simple and independent experiment that, in addition to the knowledge of the equation of state, allows for the determination of the heat capacity at constant generalized volume ${\cal V}$, $C_{\cal V}$. This experiment should be easily performed in the currently confined ultracold gases. To the best of our knowledge there are no measurements of such a heat capacity.
 It is a simple exercise to show that knowledge of the equation of state ${\cal P} = {\cal P}({\cal V}/N,T)$ and the heat capacity $C_{\cal V} = C_{\cal V}(N,{\cal V},T)$ suffices to know all the thermodynamics of a pure fluid. We need not overemphasize the fact that the quantitative features of the latter quantities are direct consequences of the interatomic interactions and of the collective interactions of the fluid.

As a corollary of our analysis we shall show that the so called ``local density approximation" (LDA) follows within the appropriate thermodynamic limit of the confining potentials. 
Our results are in agreement with 
rigorous proofs of LDA for classical systems\cite{Garrod,MarchioroI} and quantum systems\cite{MarchioroII} as well; these works have been largely overlooked in the current literature of ultracold gases, but are very relevant since they show that, under the appropriate conditions, LDA is an exact procedure. The validity of LDA is expected, and certainly widely used, because in the trapped gases the main non-uniformity appears at macroscopic length scales due to the presence of the confining external field. That is, the latter must be ``macroscopic" in order to trap a large number of particles. For macroscopic potentials this equivalently means that the energy-level separations of the external potential are much smaller than the typical atomic and collective excitations\cite{Anderson,Davies,Bradley,Greiner,Bartenstein,Thomas1,Magalhaes,Ziewerlein,Hulet,Thomas2,Shin}. Although LDA has been used in essentially all current works on ultracold gases that yield density profiles, its validity is mostly assessed on these physical grounds. 
Thus, it is reassuring to find that the virial expansion when applied to confined quantum fluids yields LDA as well. As we shall discuss, however, care must be taken when using it. That is, LDA can be directly used for the grand potential density, the entropy density and the particle density, but not so for the internal energy and other free energy densities, among other possible thermodynamic quantities. That is, while it gives a procedure to calculate thermodynamic properties of a confined inhomogeneous fluid, it does not imply that the local states of the trapped fluid are thermodynamic states of the corresponding homogenous fluid. 
 
We proceed as follows. First, we describe the general system under study - an interacting fluid confined by an external, inhomogeneous potential $V_{ext}(\vec r)$ - and we provide several specific examples to introduce the generalized volume ${\cal V}$ and the corresponding thermodynamic limit. By analyzing the system in the grand canonical ensemble, we derive the general virial expansion for the Grand Potential. We give the minimum conditions that must be obeyed by the interatomic interaction potentials for the expansion to be valid. We  discuss the physical interpretation of the generalized pressure ${\cal P}$ and show how the equation of state may be found from the knowledge of the density profile; we propose a simple experiment to measure the heat capacity at constant ${\cal V}$ in the current experiments of ultracold gases. As mentioned, LDA follows as a corollary and we discuss some of its consequences and practical uses. 

\section{Generalized volume and the thermodynamic limit.}

The system consists of $N$ identical
atoms or particles of mass $m$ with Hamiltonian
\begin{equation}
H_N = \sum_{i=1}^N \frac{\vec p_i^2}{2 m} + \sum_{i<j} u(|r_{ij}|) + 
\sum_{i=1}^N V_{ext}(\vec r_i) .\label{HN}
\end{equation}
For simplicity of presentation we have assumed additive pairwise potentials 
but, as
can be seen below, the analysis may be extended to arbitrary interatomic 
interactions. The external potential $V_{ext}(\vec r)$ is responsible 
for confining the system. To serve this purpose, it should have at least one minimum 
 and must obey that $V_{ext}(\vec r) \to \infty$ 
for $|\vec r| \to \infty$. For rigid-wall containers it is 
costumary not to write down the potential. Here, we include 
it as $V_{ext}(\vec r) = 0$ if 
$\vec r$ is within the volume $V$ enclosed by the rigid walls and 
$V_{ext}(\vec r) = \infty$ if $\vec r$ is
outside of it. Typical examples of traps of atomic gases are  
$V_{ext}(\vec r) = (1/2) m (\vec \omega \cdot \vec r)^2$ a harmonic 
potential, such as in Ref.\cite{Anderson}, and 
$V_{ext}(\vec r) = |\vec A \cdot \vec r|$
 a linear quadrupolar potential\cite{Silva};
but one can 
consider any confining potential such as a P\"oschl-Teller\cite{PT}
$V_{ext}(\vec r) = V_0/\cos(\vec \gamma \cdot \vec r)$,
 or a generic power-law potential\cite{Bagnato87}
$V_{ext}(\vec r) = \epsilon_1 |x/a|^p +  \epsilon_2 |y/b|^l + 
\epsilon_3 |z/c|^q$. We write these potentials to exemplify the appropriate 
thermodynamic variables as well as the thermodynamic limit for each case. 

To illustrate how the generalized variables appear and how the thermodynamic limit is to be taken, we shall deal here first with a classical ideal gas. Further below, we shall treat an interacting fluid, classical or quantum, and we shall verify the correctness of the identification of the variables given here. Consider, therefore, a system given by the Hamiltonian (\ref{HN}) with no interatomic interactions, i.e. $u(|r_{ij}|) \equiv 0$. Assume the system is in thermodynamic equilibrium at temperature $T$. The canonical partition function is,
\begin{equation}
Z(T,N,{\cal V}) =  \frac{1}{h^{3N} N!} \int d^{3N} p \> \int d^{3N} r \> \exp\left[{-\beta \sum_{i=1}^N \left(\frac{\vec p_i^2}{2 m} +  V_{ext}(\vec r_i)\right) }\right] ,
\end{equation}
 where $\beta = 1/kT$.  We assume the external confining potential to be of the form $V_{ext} = V_{ext}(x/l_x,y/l_y,z/l_z,\eta)$ where the quantities $l_i$ do not necessarily have units of length and $\eta$ stand for other {\it intensive} parameters that we assume to remain constant throughout. Integration of the partition function yields
\begin{equation}
Z(T,N,{\cal V}) = \frac{1}{N! \lambda_T^{3N}} \left(\zeta(\beta) {\cal V}\right)^N ,
\end{equation}  
where $\lambda_T = h/(2\pi mkT)^{1/2}$ is de Broglie thermal wavelength, ${\cal V} = l_x l_y l_z$ is the generalized volume and the function $\zeta(\beta)$ is defined by
\begin{equation}
\zeta(\beta) {\cal V} = \int \> e^{-\beta V_{ext}(\vec r)} \> d^3 r.
\end{equation}
Helmholtz free energy is found with $F = - kT \ln Z$ and, after taking the limit $N \to \infty$, yields
 \begin{equation}
 F(N,T,{\cal V}) = -NkT \left( \ln \left[\frac{{\cal V} \zeta(\beta)}{N \lambda_T^3}\right] + 1\right).
 \end{equation}
 For the free energy per particle, $F/N$, to remain finite in the thermodynamic limit, $N \to \infty$, it must be required that the ``generalized" volume diverges, i.e. ${\cal V} \to \infty$, keeping constant the ``density"  $N/{\cal V}$. As it will fully justified below,  ${\cal V}$ is an {\it extensive} thermodynamic variable. The generalized volume certainly is proportional to the actual {\it average} volume that the system occupies, $\bar V \sim \zeta(\beta) {\cal V}$. This average volume, however, is not an independent thermodynamic variable since it depends on the temperature. Moreover, it is not a correct variable since the actual volume that the system occupies is, in fact, unbounded. This can be seen by calculating the density profile which, as shown below, in this case is proportional to 
 \begin{equation}
 \rho(\vec r) \sim e^{-\beta V_{ext}(\vec r)} .
 \end{equation}
 Nevertheless,  the thermodynamic limit ${\cal V} \to \infty$ indeed implies that the volume of the system becomes arbitrarily large.
 
As it will also be justified, the generalized pressure is the intensive conjugate variable to ${\cal V}$ and may be calculated from
\begin{equation}
{\cal P} = - \left(\frac{\partial F}{\partial {\cal V}}\right)_{N,T} .
\end{equation}
For an ideal classical gas and for any confining potential, it is found that obeys ${\cal PV} = NkT$, the ideal equation of state but in the appropriate variables. 

For the particular external potentials here considered one finds, ${\cal V} = V$ for rigid walls, 
${\cal V} = 1/\omega_x \omega_y \omega_z$ for the harmonic potential, ${\cal V} = 1/A_x A_y A_z$ for the quadrupolar potential, 
${\cal V} = 1/\gamma_x \gamma_y \gamma_z$ for the P\"oschl-Teller potential
and ${\cal V} = abc$ for the generic power-law potential. Likewise, we find  $\zeta(\beta) = 1$ for rigid walls, 
$\zeta(\beta) = (2\pi /\beta m)^{3/2}$ for the harmonic potential, 
$\zeta(\beta) = 8\pi /\beta^3$ for the quadrupolar potential, 
$\zeta(\beta) = 2^3 \beta^{-(1/p + 1/l + 1/q)} \Gamma(1 + 1/p) 
\Gamma(1 + 1/l)  \Gamma(1 + 1/q)/\epsilon_1^{1/p} \epsilon_2^{1/l} 
\epsilon_3^{1/q} $ for the generic power-law potential, 
and 
$\zeta(\beta) = 4\pi \int_{-\pi/2}^{\pi /2} x^2 \exp [- \beta V_0 / \cos x] dx$ 
for the
P\"oschl-Teller potential. As we shall find below, while the role of the generalized volume is completely analogous to that of the usual volume in homogeneous systems, the thermodynamic properties of the different confined fluids show very strong variations on their temperature dependences due to the function $\zeta(\beta)$.

For purposes of presentation we have assumed that the fluids under study are effectively three-dimensional. That is, we suppose that in the three spatial dimensions the trap becomes macroscopic. It is clear that the theory can be adjusted to deal with (quasi) two- and one-dimensional systems. For this to occur, the trap must be spatially very tight  in one or two directions, which also implies that the fluid thermal excitations in those directions are smaller than the trap energy levels in the same direction; thermodynamic behavior can only exist in the remaining directions. This is the case for the very recent studies on the Berezinskii-Kosterlitz-Thouless transition in quasi 2D harmonic traps in clouds of $^{87}$Rb\cite{Zoran,Kruger} and $^{23}$Na gases\cite{Phillips}, and where the generalized area and pressure can be defined.

\section{Virial expansion for arbitrary confining potentials.}

With the identification of the generalized variables and the corresponding thermodynamic limit in hand, we now turn to the general problem of an interacting gas, classical or quantum. We extend the analysis described in the texts by Mayer and Mayer\cite{Mayer}, ter Haar\cite{Haar}
and Blatt\cite{Blatt}. Again, we assume the system is in thermodynamic equilibrium at temperature $T$ and we
analyze it in the grand canonical ensemble. We thus consider a chemical
potential $\mu$ whose value may be found by imposing a given number of
particles $N$. The grand potential is given by,
\begin{equation}
\Omega = - kT \ln \sum_{N = 0}^\infty e^{\beta \mu N} \> {\rm Tr}^\prime e^{-\beta H_N} ,
\label{Omega}
\end{equation}
where
\begin{equation}
{\rm Tr}^\prime \> e^{- \beta H_N} = \frac{1}{h^{3N} N!} \int d^{3N} r \int d^{3N} p \>
e^{- \beta H_N},
\end{equation}
if the system is classical, and
\begin{equation}
{\rm Tr}^\prime \> e^{- \beta H_N} = \frac{1}{ N!} \sum_{P} \epsilon^P \int d^{3N} r  \>
< \vec r_1, \vec r_2, \dots , \vec r_N | e^{- \beta H_N}|\vec r_{1P}, \vec r_{2P}, \dots , \vec r_{NP}>  ,
\end{equation}
if the system is quantum. The sum is over all permutations of $1,2, ... , N$ and $\epsilon = \pm 1$ for bosons or fermions.

To find the virial expansion, equation (\ref{Omega}) is first rewritten as,\cite{Mayer,Haar,Blatt}
\begin{equation}
-\beta \Omega = \sum_{n=1}^\infty \> e^{ \beta \mu n} \> \frac{1}{n!} \> I_n, \label{Ome-vir}
\end{equation}
where the functions $I_n$ are given by,
\begin{eqnarray}
I_n = \left\{
\begin{array}{c} \frac{1}{h^{3n}} \int d^{3n} r \int d^{3n} p \> U_n(\vec r_1, \vec p_1; \dots; \vec r_n, \vec p_n) \\ \\
\int d^{3n} r \>U_n(\vec r_1, \dots, \vec r_n)
\end{array} \right.
\end{eqnarray}
for classical and quantum systems, respectively, and where the Ursell functions are given by the hierarchy: first order,
\begin{equation}
U_1(1) = W_1(1) ,
\end{equation}
second order,
\begin{equation}
U_2(1,2) = W_2(1,2) - U_1(1) U_1(2) ,
\end{equation}
third order,
\begin{eqnarray}
U_3(1,2,3) &=&  W_3(1,2,3) - U_1(1) U_2(2,3) -U_1(2) U_2(1,3) - U_1(3) U_2(1,2)- \nonumber \\
&&U_1(1)U_1(2)U_1(3) .
\end{eqnarray}
and so on, and 
\begin{eqnarray}
W_n(1,2, ..., n) = \left\{
\begin{array}{c}  e^{- \beta H_n} \\ \\
 \sum_{P} \epsilon^P  \>
< \vec r_1, \vec r_2, \dots , \vec r_n | e^{- \beta H_n}|\vec r_{1P}, \vec r_{2P}, \dots , \vec r_{nP}>
\end{array} \right.
\end{eqnarray}
for classical and quantum systems, respectively.

The problem of the virial expansion reduces to find the value of each contribution $I_n$ in the thermodynamic limit, taking into account the interactions among the atoms or molecules. This is what we do now for a general confining external potential $V_{ext}(\vec r)$. We proceed by systematically calculating $I_n$ order by order and then generalize it to $I_n$. We did so from $I_1$ to $I_4$. Since the calculations are quite lengthy, though straightforward, we explicitly present in the Appendix the case $I_2$ only. Next we discuss the results.

The calculation of $I_1$ is very simple and it turns out that, in the thermodynamic limit, the classical and quantum cases give the same result:

\begin{equation}
I_1 = \left\{ \begin{array}{c} \frac{1}{h^3} \int d^3 p \int d^3 r \>  e^{-\beta H_1} \\ \\
\int d^3 r < \vec r | e^{- \beta H_1} | \vec r > \end{array} \right.
\end{equation}
where the one-particle Hamiltonian $H_1$ is given by 
\begin{equation}
H_1 = {\vec p^2 \over 2m} + V_{ext}(\vec r) .
\end{equation}
In the limit, one finds,
 \begin{equation}
I_1 = {1 \over \lambda_T^3} \zeta(\beta) {\cal V}  \label{I1}
 \end{equation}
 for both classical and quantum cases.

In the Appendix we show the explicit calculation of $I_2$. That analysis suffices to see how to find $I_n$. The key is in the separation of center of mass and relative coordinates. This change of variables is generally, $
\vec R = \frac{1}{n} (\vec r_1 + \vec r_2 + \dots + \vec r_n)$, $\vec r^{(1)} = \vec r_1 - \vec r_2$,
$\vec r^{(2)} = \vec r_2 - \vec r_3$, $\dots$, $\vec r^{(n-1)} = \vec r_{n-1} - \vec r_n$, with their canonical conjugate momenta.  

In both the classical and quantum cases, the main assumption is the same, namely, that one must take the thermodynamic limit ${\cal V} \to \infty$. This  allows to make the approximation,
\begin{equation}
V_{ext}(\vec r_1) + V_{ext}(\vec r_2) + \dots + V_{ext}(\vec r_n) \approx n V_{ext}(\vec R) ,\label{terlimn}
\end{equation}
where $(\vec r_1, \vec r_2, \dots, \vec r_n)$ are to be given in terms of the variables $(\vec R,  \vec r^{(1)} ,\vec r^{(2)} ,\dots, \vec r^{(n-1)} )$ by the above transformation. This approximation separates the center of mass motion from the relative ones. The former is always quasiclassical and its contribution to $I_n$ is proportional to $\zeta(n\beta) {\cal V}/\lambda_T^{3n}$, while the contribution from the relative coordinates yields the virial coefficients $b_n$; these can be classical or quantum, but they are the {\it same} for any external potential. That is, we find
\begin{equation}
I_n = \frac{\cal V}{\lambda_T^{3n}} \zeta(n\beta) \>b_n .\label{In}
\end{equation}
The validity of the above procedure, in the classical case, reduces to require that 
the intermolecular potential must vanish for lengths $r \gg \sigma$, with $\sigma$ the range of such a potential. Additionally, the interaction must be ``short-range", namely, decaying faster than $1/r^3$, otherwise the virial coefficients $b_n$ do not exist\cite{Garrod,Mayer}. In the quantum case and for high temperatures, the range of the relative variables $\vec r$ is bounded due to presence of the potential $u(r)$ and the validity has the same limitations as in the the classical case. At low temperatures the bound is set up by either the thermal de Broglie wavelength or the scattering length $a$\cite{LLQM}. If the gas behaves as an ideal one, the relevant length is de Broglie wavelength. In any case, as long as the relative coordinates remain bounded by a {\it finite} quantity, however large, one can take the limit of very large volumes, ${\cal V} \to \infty$, and implement the thermodynamic limit just as in the classical case. 

Summarizing, we find that in the thermodynamic limit the grand potential can be written in general as,
\begin{equation}
\Omega = - kT {\cal V} \sum_{n=1}^\infty \frac{e^{\beta n \mu}}{n!} \frac{\zeta(n \beta)}{\lambda_T^{3n}} b_n(T)
\label{Ob}
\end{equation}
where quantum or classical virial coefficients should be used; in both cases $b_1 = 1$. Expression (\ref{Ob}) for the grand potential is one of  the main results of this article. It is
the virial expansion for arbitrary confining potentials. 
The number of particles $N$ and the entropy $S$ can be calculated from (minus) the 
partial derivatives of $\Omega$ with respect to $\mu$ and $T$ respectively.  
$\Omega$, $N$ and $S$ are found to be homogeneous first order functions
of ${\cal V}$, and  this implies that ${\cal V}$ must be an extensive variable and justifies the thermodynamic
limit as used above. From the expression $\Omega = - {\cal PV}$,  the generalized pressure is read off (\ref{Ob}).

Note that the most important difference of the grand potential between a given arbitrary external potential and the homogeneous case is the function $\zeta({\beta})$ rather than the generalized volume ${\cal V}$. The latter enters in the same way for any potential, including the rigid-walls case; that is, it gives rise to the intensive quantities formed between the extensive variables $N$, $S$, $E$ etc. and ${\cal V}$, and that remain finite in the thermodynamic limit, i.e. $N/{\cal V}$, $S/{\cal V}$, $E/{\cal V}$, etc. However, as it is well known from calculations in external potentials, the temperature dependence of the thermodynamic variables is very different and unique for each external potential. This difference is contained in the function $\zeta({\beta})$. This will be exemplified in the next section.

To illustrate the use of Eq.(\ref{Ob}), we apply it to an {\it ideal} quantum gas. From the analysis in the Appendix and their corresponding value for third and fourth orders, one finds that the virial coefficients are given by
\begin{equation}
b_n^{(0)} = \epsilon^{n+1} \> \frac{n!}{n^{5/2}} \> \lambda_T^{3(n-1)} .
\end{equation}
The grand potential for an ideal quantum gas can, thus, be written as
\begin{equation}
-\beta \Omega =  \frac{\cal V}{\lambda_T^3} \sum_{n=1}^\infty \> e^{n \beta \mu} \> \zeta(n\beta)\>\frac{\epsilon^{n+1}}{n^{5/2}} . \label{omeiq}
\end{equation}
This formula can be directly compared with the corresponding ones for, say, the rigid walls potential ${\cal V} = V$ and $\zeta(n \beta) = 1$, or the harmonic potential ${\cal V} = 1/\omega^3$ and $\zeta(n\beta) =   (2\pi kT/nm)^{3/2}$. The ``textbook"  formulae for these potentials are,\cite{Pethick}
\begin{equation}
-\beta \Omega  = \frac{V}{\lambda_T^3} \frac{1}{\Gamma(5/2)}\int_0^\infty \frac{x^{3/2} dx}
{e^{x-\beta\mu} - \epsilon} 
\end{equation}
for rigid walls, and
\begin{equation}
-\beta \Omega  = \left(\frac{kT}{\hbar \omega}\right)^3 \frac{1}{\Gamma(4)}\int_0^\infty \frac{x^{3} dx}
{e^{x-\beta\mu} - \epsilon} 
\end{equation}
for a 3D isotropic harmonic potential. Expansion of the integrals of these last two equations in powers of $e^{\beta \mu}$ yield the virial expansion, Eq.(\ref{omeiq}).

\section{Physical meaning of the generalized pressure. Equation of state and heat capacity.}

In addition to the previous discussions, we want to emphasize several points that
we consider to be of importance for both the understanding of the present results and their practical 
applications. 

First,
we  recall a result previously set forth by one of the 
authors\cite{RR1,RR2,RR3}. This is the fact that since 
${\cal V}$ is an {\it extensive} generalized volume, there exists 
an {\it intensive} generalized pressure ${\cal P}$, 
conjugate to ${\cal V}$, given by
\begin{equation}
{\cal P} = - \left( \frac{\partial \Omega}{\partial {\cal V}} \right)_{T,\mu} .
\end{equation}  
and, therefore, $\Omega = - {\cal P V}$. 
A relevant consequence is that
Euler relationship is
$E = TS - {\cal P}{\cal V} + \mu N$ for {\it any} external potential in the limit here discussed.
Thus, ${\cal P}$ being a {\it bona-fide} thermodynamic 
variable allows us to calculate, for instance,
the Gibbs potential $G = G(T,N,{\cal P})$ or 
the entalphy $H = H(S,N,{\cal P})$ by 
appropriate Legendre transforms. The identification of the generalized 
pressure also yields
the correct equation of state of the fluid; this is the 
relationship ${\cal P} = {\cal P}({\cal V}/N, T)$. 

Second, 
the identification of ${\cal P}$ is not only a formal one but it
has a clear physical interpretation as the variable responsible for 
mechanical equilibrium. It must be
recognized that the usual hydrostatic pressure is not the appropriate 
thermodynamic variable
in an inhomogeneous fluid since it is a local variable. In general, however, mechanical equilibrium of an inhomogeneous fluid is given by Pascal law,
\begin{equation}
\nabla \cdot \tilde P(\vec r) = - \rho(\vec r) \nabla V_{ext}(\vec r) ,\label{Pascal}
\end{equation}
where $\tilde P(\vec r)$ is the pressure tensor of the fluid and $\rho(\vec r)$ is the local particle density, namely, the density profile. As we shall see below, barring phase-separated states within the confined fluid, one expects  the 
pressure tensor to be a local quantity, $\tilde P(\vec r) = p(\vec r) \tilde 1$,
where $\tilde 1$ is the unit tensor and $p(\vec r)$ the local hydrostatic
pressure.
Integration of the virial of the external force, namely, of 
$- \vec r \cdot \left( \rho(\vec r) \nabla V_{ext}(\vec r) \right)$, yields
\begin{eqnarray}
\int {\rm Tr} \tilde P \> d^3 r &=&  \int d^3 r \> 
\rho(\vec r) \> \vec r \cdot \nabla V_{ext}(\vec r) \nonumber \\
&=&  \left< \sum_{i=1}^N \vec r_i \cdot \nabla_i V_{ext}(\vec r_i) \right> , \label{tr}
\end{eqnarray}
where in the second line we have identified the average of the virial of the external force. This average can be performed in any ensemble. Let us choose the canonical ensemble. The partition function is
\begin{equation}
Z = {\rm Tr} \> e^{-\beta H_N}
\end{equation}
with $H_N$ the Hamiltonian of the interacting $N$ particles confined by $V_{ext}(\vec r)$, as given by Eq.(\ref{HN}). Recalling that Helmoltz free energy is $F = -kT \ln Z$, one finds $F = F(N,T,{\cal V})$, and by a simple calculation,
\begin{eqnarray}
{\cal P}& =&   - \left(\frac{\partial F}{\partial {\cal V}}\right)_{T,N} \nonumber \\
&=& \frac{1}{3 \cal V} \left< \sum_{i=1}^N \vec r_i \cdot \nabla_i V_{ext}(\vec r_i) \right> .
 \end{eqnarray}
This result has a two-fold relevance. Comparing with Eq.(\ref{tr}), one finds the physical significance of the generalized pressure, 
\begin{equation}
{\cal P V} = \frac{1}{3} \int {\rm Tr} \tilde P \> d^3 r .
\end{equation}
It is simply an intensive global property of the pressure tensor, the quantity that bears the information of mechanical equilibrium in the body. An alternative form is clearly,
\begin{equation}
{\cal P}{\cal V} = \frac{1}{3} \int d^3 r \> 
\rho(\vec r) \> \vec r \cdot \nabla V_{ext}(\vec r)  \label{PV}
\end{equation}
which may be seen as the generalization of the celebrated theorem found by Clausius in the middle of the XIX century for homogenous systems: The (average) virial of the  external force equals $-3 {\cal PV}$. It follows as well from the present discussion that 
the reversible work performed on or by the system after a change of the external potential is given by,
\begin{equation}
d W = - {\cal P} \> d {\cal V} .
\end{equation} 

Further, in addition to the interpretation of the generalized pressure, we point out that Eq.(\ref{PV}) is a remarkable formula 
since it gives rise, with the sole knowledge of the density profile
$\rho(\vec r)$, to a direct route for the
calculation of the generalized pressure and, hence, to the
equation of state of the fluid, ${\cal P} = {\cal P} ({\cal V}/N,T)$. We recall that  
the density profile may be known from exact or approximate calculations, or directly
from 
experiments. It is somewhat puzzling to realize that expression (\ref{PV}) yields only an identity for the rigid-wall case and does not give a calculational tool for the hydrostatic pressure. The latter needs the knowledge, at least for pairwise interatomic interactions, of the two-body density correlation function\cite{RW}. Here, we find that for inhomogeneous systems knowledge of the one-body density suffices. This was the procedure used in 
Ref.\cite{RR1} to obtain the equation of state of
a degenerate interacting confined Bose gas within the Hartree-Fock approximation.

For completeness of our presentation, we write down the first few terms  of the so-called virial expansion of the equation of state ${\cal P} = {\cal P} (N/{\cal V},T)$. This can be done by finding $N = N(\mu,T,{\cal V})$ from Eq.(\ref{Ob}) and inverting it term by term to yield $\mu = \mu(N/{\cal V},T)$, then, substituting the result into ${\cal P} = {\cal P}(\mu,T)$:
\begin{equation}
{\cal P}(\frac{N}{\cal V},T) = \frac{N}{\cal V} kT \left[1 - \frac{1}{2} \frac{\zeta(2\beta)}{\zeta^2(\beta)} b_2(T) \frac{N}{\cal V}
+ \left(\frac{\zeta^2(2\beta)}{\zeta^4(\beta)} b_2^2(T) - \frac{2}{3} \frac{\zeta(3\beta)}{\zeta^3(\beta)}b_3(T)\right)
\left(\frac{N}{\cal V}\right)^2 + \dots \right] .
\end{equation}
Once again, we remark that the functions $\zeta(\beta)$ make all the difference. Since in some instances one can refer the calculation of the virial coefficients to a diagramatic expansion\cite{Mayer}, one finds that the diagrams sum up differently for different potentials. We also recall that this type of virial expansion was used in Ref.\cite{Silva} to fit experimental data from a gas of Sodium atoms in a quadrupolar potential.

	As a further instance of the relevance of these variables, we mention here that there has been a lot of interest recently on the behavior of fermi gases ($^{40}$K and $^{6}$Li) near the unitarity limit where the scattering length diverges, because it appears that thermodynamics becomes universal, i.e. independent of the interatomic interactions; see e.g. Refs.\cite{Thomas1,Ziewerlein,Hulet,Thomas2,Shin,Ho,Drummond,Thomas3}. In this region, the gas behaves as an ideal one in the sense that it obeys the ideal virial relationship $E = 2N <V_{ext}>$ for a harmonic trap, with $E$ the internal energy. The relevance to the present work is that $N<V_{ext}>$ for a harmonic trap is $3/2 {\cal PV}$, whether the gas behaves as ideal or not, see Eq.(\ref{PV}) and Ref.\cite{RR1}. That is, the quantity that has been {\it measured} in Refs.\cite{Thomas1,Thomas2,Drummond,Thomas3} is precisely the generalized pressure for the harmonic trap. And indeed, if the universality hipothesis is correct\cite{Ho}, using the virial theorem for ideal gases for arbitrary potentials, yields the following relation that should be obeyed in the unitarity region,
\begin{equation}
E = < \sum_{i=1}^N V_{ext}(\vec r_i)> + \frac{3}{2}{\cal PV} .
\end{equation}
Away from the unitarity limit, this equation is no longer valid but the measurement of ${\cal P}$, and of the heat capacity as we describe below, can be performed to obtain the thermodynamics of those states.

The equation of state does not suffice to know all the thermodynamic properties of a one-component fluid. One needs a second relationship such as the heat capacity at constant generalized volume $C_{\cal V} = C_{\cal V}(N,{\cal V},T)$. We propose now a very simple measurement to find out such a heat capacity in the current ultracold fluids. This consists of an adiabatic compression or expansion: we note that the ultracold trapped gases are actually isolated and confined by magnetic or optical traps\cite{Anderson,Davies,Bradley,Greiner,Bartenstein,Thomas1,Magalhaes,Ziewerlein,Hulet,Thomas2,Shin} and, therefore, a slow change of the confining potential, namely, of the generalized volume, will give rise to an increase or decrease of the temperature, depending on whether the fluid is compressed or expanded. This procedure is adiabatic. Since the generalized volume and temperature are measured in the current experiments, the quantity $(\partial {\cal V}/\partial T)_{N,S}$ can thus also be measured. Then, from a thermodynamic relationship one can evaluate the corresponding heat capacity,
\begin{equation}
C_{\cal V} = - T \left(\frac{ \partial {\cal P}}{\partial T}\right)_{N/{\cal V}} \left(\frac{\partial {\cal V}}{\partial T}\right)_{S,N} ,
\end{equation}
where previous knowledge of the equation of state is needed for the calculation of the second factor on the right hand side. Note that the measurements of the equation of state and of the heat capacity correspond to two different sets of experiments.

\section{A comment on the ``local density approximation".}

The exactness, or validity, of the ``local density approximation" 
follows right away from
the corresponding expressions for $\Omega$, $N$ and $S$, as given by Eq.(\ref{Ob}). Consider the 
rigid-wall external potential, ${\cal V} = V$ and $\zeta(n \beta) = 1$. 
We can thus define the grand potential per unit  volume $\omega(\mu,T) = \Omega/V$, 
the number of particles per unit volume (particle density) $\rho(\mu,T) = N/V$ and
the entropy per unit volume $s(\mu,T) = S/V$. We now consider the same system but confined by an external potential $V_{ext}(\vec r)$. We can find its thermodynamic properties by implementing the ``local density approximation'': take $\omega$, $\rho$ and $s$ of the homogenous case and make those functions 
per unit volume to be their  ``local" densities 
$\omega(\vec r)$, $\rho(\vec r)$ and $s(\vec r)$, in the presence 
of the given external potential, by replacing the 
chemical potential $\mu$ by the ``local" chemical potential 
$\mu_{local}(\vec r) = \mu - V_{ext}(\vec r)$. It turns out that integration of $\omega(\vec r)$, 
$\rho(\vec r)$ and $s(\vec r)$ over
all space yield the {\it exact} expansions for $\Omega$, $N$ 
and $S$, in the presence of $V_{ext}(\vec r)$, as given by Eq.(\ref{Ob}) and its
derivatives. That is, one finds that LDA procedure gives rise to exact results. We recall that the validity of LDA for classical and quantum systems in this limit was rigorously proved in Refs.\cite{Garrod,MarchioroI} and \cite{MarchioroII}, respectively.

The above description does show that in the thermodynamic limit the system is locally homogenous and that ``locally" actually means in length scales large  compared with those of interatomic interactions. It is in this latter connection that LDA is largely used without the need of further justification. There is a warning, however, that must be raised when using LDA. It may appear that if one is able to find {\it any} thermodynamic variable $q$ for a homogenous system and express it in terms of the chemical potential $\mu$ and temperature $T$, namely $q = q(\mu,T)$, its local counterpart when in the presence of an external potential $V_{ext}(\vec r)$ is simply $q(\vec r) = q(\mu_{local}(\vec r),T)$. This, in general, is incorrect; its is strictly justified for $\Omega/V$, $N/V$ and $S/V$ only . It is incorrect, for instance, for the internal energy and other free energies, except $\Omega$, as well as for other functions such as heat capacities. This statement can be verified by using expression (\ref{Ob}) for the Grand Potential of a confined fluid.
Thus, the fact that the system is locally homogeneous does not imply that the local states of the confined fluid are thermodynamic states of the corresponding homogeneous system, i.e. $q_{local}(\vec r) \ne q(\mu_{local}(\vec r),T)$ in general.

\section{Final Remarks}

As a final comment, we emphasize that the use of  the generalized thermodynamic variables here presented should lead to correct results by simply following the rules of thermodynamics, without necessarily resorting to a local picture. It is clear to us that for this to be useful one needs, first, to change the usual intuition on volume and hydrostatic pressure to their generalized counterparts and, second, to provide examples where these variables lead to new insights. There are already few studies where the latter has been put to use in approximated calculations of equations of state\cite{RR1,RR3} and there already incipient experiments that have made use of them\cite{Silva,Henn,Henn2}. Our contention is that the measurement of the equation of state and the heat capacity in terms of the generalized thermodynamic variables should offer a complementary and useful tool for the analysis of ultracold trapped fluids.

Although much has been learned with use of LDA, we have indicated that care must be taken when using it. This should be more notorious when dealing with phase-separated fluids where it is not clear if LDA suffices for their description since the interfacial widths of the phase boundaries are expected to be of the order of the range of the intermolecular 
interactions\cite{RW}.  This situation appears to be the case for the states found recently in trapped gases of $^6$Li atoms\cite{Ziewerlein,Hulet,Shin}, where it is found that the confined fluid phase-separates into a superfluid and a normal paramagnetic gas, showing clearly an interfacial phase boundary. There are differing theoretical studies of whether LDA should be enough or if ``surface tension" terms should be included\cite{DeSilva,Gubbels,Chien}. In general, for such inhomogeneous states, one should not expect a local picture to be valid across the interface; the thermodynamic 
potentials are indeed expected to
be non-local on the density profiles.

\begin{acknowledgments}
Work supported by UNAM DGAPA IN-114308. We thank an anonymous referee for bringing to our attention Refs.\cite{Garrod}, \cite{MarchioroI} and \cite{MarchioroII}. 
\end{acknowledgments}

\appendix*

\section{Derivation of $I_2$}

In this Appendix we provide the derivation of the second order contribution to the virial expansion.

{\it Classical case.} From Section III we find that $I_2$ in the classical case is given by
\begin{eqnarray}
I_2 &=& {1 \over h^6}\int d^3 p_1 \int d^3 r_1\int d^3 p_2 \int d^3 r_2 \> e^{- \beta H_2} -
\nonumber \\
&&{1 \over h^6} \left(\int d^3 p_1 \int d^3 r_1  \> e^{- \beta H_1} \right)
\left( \int d^3 p_2 \int d^3 r_2\> e^{- \beta H_1}  \right)
\end{eqnarray}
with $H_2$ given by eq.(\ref{HN}) for $N = 2$.
We make the change of variables to center of mass and relative coordinates, 
\begin{equation}
\vec R = (\vec r_1 + \vec r_2)/2 \>\>\>{\rm and} \>\>\>
\vec r = \vec r_1 - \vec r_2, \label{ch2}
\end{equation}
 with their canonical momenta $\vec P$ and $\vec p$. Thus,
\begin{eqnarray}
I_2 &=& \frac{1}{h^6} \int d^3 P \int d^3 R \int d^3 p \int d^3 r \> e^{- \beta H_{cm}^{(2)}} \> e^{-\beta {\cal H}_2}-
\nonumber \\
&& {1 \over \lambda_T^6} \int d^3 R  \int d^3 r e^{-\beta (V_{ext}(\vec R + {\vec r \over 2}) + V_{ext}(\vec R - {\vec r \over 2}))}
\end{eqnarray}
where the center of mass Hamiltonian is 
\begin{equation}
H_{cm}^{(2)} = {\vec P^2 \over 2 (2m)} + V_{ext}(\vec R + {\vec r \over 2}) + V_{ext}(\vec R - {\vec r \over 2})
\end{equation}
and the 2-particle relative-coordinates one is,
\begin{equation}
{\cal H}_2 = {\vec p^2 \over 2 (m/2)} + u(r). \label{H2rel}
\end{equation}
The integrals over the momenta yield $1/\lambda_T^6$.
Notice, however, that  the integrals over $\vec R$ depend strongly on $\vec r$, even for particles confined in a rigid vessel of volume $V$. In such a case, the boundary terms couple the integrals.   
We rewrite $I_2$ as follows,
\begin{equation}
I_2 = {1 \over \lambda_T^6} \int d^3 R  \int d^3 r e^{-\beta (V_{ext}(\vec R + {\vec r \over 2}) + V_{ext}(\vec R - {\vec r \over 2}))} f(r)
\end{equation}
where we have introduced the Mayer function,
$f(r) = e^{-\beta u(r)} -1.$ 
Here is the important step: the function $f(r)$ vanishes for distances $r$ longer than the range of the intermolecular potential $\sigma$, i.e. the value of
$r$ is bounded, $r \le \sigma$. Hence, in the thermodynamic limit, in which the external potential becomes ``shallower" and ``shallower" and the volume of the system larger and larger, we can set,
\begin{equation}
V_{ext}(\vec R + {\vec r \over 2}) + V_{ext}(\vec R - {\vec r \over 2}) \approx 2 V_{ext}(\vec R) .\label{terlim}
\end{equation}
This is the thermodynamic limit. Therefore,
\begin{equation}
I_2 = {1 \over \lambda_T^6} \int d^3 R e^{-2 \beta V_{ext}(\vec R)}  \int d^3 r f(r) = 
 {{\cal V} \over  \lambda_T^6} \zeta(2 \beta) b_2 , \label{cb2}
 \end{equation}
 where we have identified the classical second virial coefficient $b_2$. Thus, the well-known bound is that the intermolecular potential must vanish for lengths $r \gg \sigma$. Clearly, it must be ``short-range" interaction (faster than $1/r^3$), otherwise $b_2$ does not exist. 
  
{\it Quantum case.} Again, from the expressions in Section III, one finds,
\begin{eqnarray} 
I_2 &=& \int d^3 r_1 \int d^3 r_2 \left( < \vec r_1, \vec r_2 | e^{-\beta H_2} |  \vec r_1, \vec r_2 > +
\epsilon < \vec r_1, \vec r_2 | e^{-\beta H_2} |  \vec r_2, \vec r_1 >  \right) - \nonumber \\
&&  \int d^3 r_1 < \vec r_1| e^{-\beta H_1} |  \vec r_1 > \> 
\int d^3 r_2 <  \vec r_2 | e^{-\beta H_1} |   \vec r_2 >  
\end{eqnarray}
where $\epsilon = \pm 1$ for bosons or fermions. The derivation follows essentially the same steps as in the classical case. First, one performs the same change of variables as in Eq.(\ref{ch2}) to rewrite $I_2$. Then, the thermodynamic limit may be taken by separating the center of mass motion from the relative one. It can be realized that for high temperatures the range of the variable $\vec r$ is bounded due to presence of the potential $u(r)$. At low temperatures the bound is set up by either the thermal de Broglie wavelength or the scattering length $a$.  Therefore, as long as $r$ remains bounded by a {\it finite} quantity, however large, one can take the limit of very large volumes, ${\cal V} \to \infty$, and implement the thermodynamic limit just as in the classical case. One finds,
\begin{equation}
I_2 =  {{\cal V} \over  \lambda_T^6} \zeta(2 \beta)  b_2  \label{qb2},
 \end{equation}
formally identical with its classical counterpart, formula (\ref{cb2}), but with the {\it quantum} second virial coefficient: 
\begin{eqnarray}
b_2 &=& 2^{3/2} \lambda_T^3 \int d^3 r \left(
< \vec r | e^{-\beta {\cal H}_2} |\vec r> + \right. \nonumber \\
&&\left. \epsilon < \vec r | e^{-\beta {\cal H}_2 }|-\vec r> -
< \vec r | e^{-\beta p^2/2(m/2)}|\vec r> \right) , \label{qb2a}
\end{eqnarray}
where  ${\cal H}_2$ is given by  Eq.({\ref{H2rel}). As a rule, in the thermodynamic limit the center of mass motion is always quasiclassical\cite{LL}. The expression for the quantum second virial coefficient above can be seen to be the correct one by comparing, for instance, with the expression given in Ref.\cite{LL}. For slow collisions, the relevant ones for ultracold gases, $b_2$ depends on the scattering length and this may become quite large near a Feschbach or potential resonance. The formulae here derived may then be not applicable very near such a point, called the unitarity limit, but as it has been shown\cite{Ho} this may be expected since in such a limit  the system behaves as if near a critical point. We add that the description of the scattering al low energies near resonances is valid for interatomic potentials $u(r)$ that decay at least as $1/r^3$\cite{LLQM}.

To end this part, we find illustrative to calculate $b_2$ for an ideal quantum gas, i.e. for $u(r) = 0$. One finds the so-called ``exchange" contribution to the second virial coefficient:
\begin{equation}
b_2^{(0)} =  \epsilon \>  \frac{1}{2^{3/2}} \> \lambda_T^3.
 \label{b20}
\end{equation}

\end{document}